\documentclass[preprint,amsmath,amssymb,superscriptaddress,nobalancelastpage,aps,longbibliography,showpacs]{revtex4-2}

\usepackage{braket}
\usepackage{varioref}
\usepackage{xr-hyper}
\usepackage{xcolor}
\usepackage{nicefrac}
\usepackage{xfrac}
\usepackage{hyperref}
\hypersetup{colorlinks,linkcolor=blue,urlcolor=blue,citecolor=orange}
\usepackage{ulem}
\usepackage{siunitx}
\usepackage{graphicx}
\usepackage{dcolumn}
\usepackage{bm}

\usepackage{color}  

\newcommand{\blue}[1] {\textcolor{blue}{#1}}
\newcommand{\cut}[1] {\textcolor{green}{[----]}}

\begin{document}

\title{Unified Description of Charge Density Waves \\ in Electron- and Hole-doped Cuprate Superconductors}

\author{Jaewon Choi}
\email{jaewon.choi@diamond.ac.uk}
\affiliation{Diamond Light Source, Harwell Campus, Didcot OX11 0DE, United Kingdom}
\affiliation{Department of Physics, Korea Advanced Institute of Science and Technology (KAIST), 291 Daehak-ro, Daejeon 34141, Republic of Korea}

\author{Sijia Tu}
\affiliation{Beijing National Laboratory for Condensed Matter Physics and Institute of Physics, Chinese Academy of Sciences, Beijing 100190, China}
\affiliation{University of Chinese Academy of Sciences, Beijing 100049, China}

\author{Abhishek Nag}
\affiliation{Diamond Light Source, Harwell Campus, Didcot OX11 0DE, United Kingdom}
\affiliation{Laboratory for Non-linear Optics, Paul Scherrer Institut, Villigen CH-5232, Switzerland}

\author{Charles C. Tam}
\affiliation{Diamond Light Source, Harwell Campus, Didcot OX11 0DE, United Kingdom}
\affiliation{H. H. Wills Physics Laboratory, University of Bristol, Bristol BS8 1TL, United Kingdom}

\author{Sahil Tippireddy}
\affiliation{Diamond Light Source, Harwell Campus, Didcot OX11 0DE, United Kingdom}

\author{Stefano Agrestini}
\affiliation{Diamond Light Source, Harwell Campus, Didcot OX11 0DE, United Kingdom}

\author{Zefeng Lin}
\affiliation{Beijing National Laboratory for Condensed Matter Physics and Institute of Physics, Chinese Academy of Sciences, Beijing 100190, China}
\affiliation{University of Chinese Academy of Sciences, Beijing 100049, China}

\author{Mirian Garc\'{i}a-Fern\'{a}ndez}
\affiliation{Diamond Light Source, Harwell Campus, Didcot OX11 0DE, United Kingdom}

\author{Kui Jin}
\affiliation{Beijing National Laboratory for Condensed Matter Physics and Institute of Physics, Chinese Academy of Sciences, Beijing 100190, China}
\affiliation{University of Chinese Academy of Sciences, Beijing 100049, China}

\author{Ke-Jin Zhou}
\email{kejin.zhou@diamond.ac.uk}
\affiliation{Diamond Light Source, Harwell Campus, Didcot OX11 0DE, United Kingdom}
\date{\today}

\begin{abstract}
High-temperature cuprates superconductors are characterised by the complex interplay between superconductivity (SC) and charge density wave (CDW) in the context of intertwined competing orders. In contrast to abundant studies for hole-doped cuprates, the exact nature of CDW and its relationship to SC was much less explored in electron-doped counterparts. Here, we performed resonant inelastic x-ray scattering (RIXS) experiments to investigate the relationship between CDW and SC in electron-doped La$_{2-x}$Ce$_x$CuO$_4$. The short-range CDW order with a correlation length $\sim35$~\AA~was found in a wide range of temperature and doping concentration. Near the optimal doping, the CDW order is weakened inside the SC phase, implying an intimate relationship between the two orders. This interplay has been commonly reported in hole-doped La-based cuprates near the optimal doping. We reconciled the diverging behaviour of CDW across the superconducting phase in various cuprate materials by introducing the CDW correlation length as a key parameter. Our study paves the way for establishing a unified picture to describe the phenomenology of CDW and its relationship with SC in the cuprate family.

\end{abstract}

\maketitle

\section{Introduction}

Unconventional superconductivity (SC) emerges when charge carriers, either electrons or holes, are introduced into antiferromagnetic (AFM) parent compounds \cite{Dagotto2005}. In copper-oxide (cuprate) superconductors, exotic broken-symmetry phases such as pseudogap, nematicity, and various density waves coexist in the vicinity of SC. The phenomenology of these electronic states is expected to be symmetric upon doping either electrons or holes, when the parent compounds are described by simplest theoretical models such as effective single-band Hubbard model \cite{Armitage2010}. However, in reality, the electron- and hole-doped regimes in the cuprate phase diagram are not entirely symmetric \cite{Greene2020} (See Fig.~\ref{fig1}a), despite the presence of SC \cite{Takagi1989}, anti-ferromagnetism (AFM), spin \cite{WSLee2014,XTLi2024} and charge excitation \cite{daSilvaNeto2018,daSilvaNeto2015,daSilvaNeto2016,Jang2017,Hepting2018,Lin2020}. For example, the AFM phase in the electron-doped cuprates is more robust and extends to the SC phase boundary comparing to the hole-doped cuprates. The SC phase in the electron-doped cuprates is confined into a narrower region of the phase diagram \cite{Takagi1989}. Unlike the broad appearance in the phase diagram of hole-doped cuprates, the pseudogap is absent in electron-doped cuprates~\cite{Greene2020}. These similarities and differences may hold a key to distinguish the essential ingredients for the emergence of SC phase. 

The interplay between different broken-symmetry phases  often leads to the formation of exotic intertwined orders  in hole-doped cuprates, beyond simple coexistence \cite{Kivelson2003}. In particular, a charge-density-wave (CDW), a periodic modulation of valence electrons commonly discovered in virtually all cuprate families \cite{Tranquada96,Zimmermann1998,Hoffman2002,Howald2003,Hanaguri2004,Abbamonte2005,DoironLeyraud2007,LeBoeuf2007,Wu2011,Wise2008,Parker2010,Ghiringhelli2012,Chang2012,Neto2014,Comin2014,Wu2014}, is intimately linked to SC and, sometimes, spin-density wave (SDW), thus plays a central role in the underlying mechanism of superconductivity. For example, CDW competes with SC order in La- and Y-based cuprates \cite{Chang2012,Chang2016,Choi2020} near 1/8 doping concentration, evidenced by suppression of its intensity below $T_c$ accompanied with a dip in SC phase \cite{Comin2016}. It is further proposed to form a pair-density-wave (PDW) state where SC is spatially modulated due to the interplay with CDW and SDW \cite{Kivelson2003,Wen2019}. In Bi-based cuprates Bi$_2$Sr$_2$CaCu$_2$O$_{8+\delta}$ (Bi-2212), the reduction of CDW intensity inside SC phase was interpreted beyond a simple competition picture as evidence of quantum fluctuations driven by Fano interference, associated with the interplay between CDW fluctuations and phonon~\cite{WSLee2019,Lu2022}. In comparison, the CDW intensity does not present obvious change upon entering the SC phase in Bi$_2$Sr$_{2-x}$La$_{x}$CuO$_{6+\delta}$ (Bi-2201)~\cite{Comin2014,Li2020,Choi2024}. The seemingly contradicting behaviour of CDW in various hole-doped cuprates casts doubts on whether there is a universal relationship between CDW and SC.

In contrast to abundant experimental results provided for the hole-doped regime, the relationship between SC and CDW orders is even less clear in electron-doped cuprates. The existence of CDW order was confirmed by resonant x-ray scattering (RXS) in Nd$_{2-x}$Ce$_{x}$CuO$_4$ (NCCO) and La$_{2-x}$Ce$_{x}$CuO$_4$ (LCCO). However, these studies do not suggest any interaction between CDW and SC \cite{daSilvaNeto2018,daSilvaNeto2015,daSilvaNeto2016,Jang2017}. The above diverging experimental results raise several questions: What is the intrinsic relationship between SC and CDW in electron-doped cuprates? Is the absence of the interaction a common feature shared among electron-doped cuprates or only specific to certain materials? Answers to these questions would help elucidate a more fundamental question of whether the intimate interplay between SC and CDW is a generic phenomenology in both hole- and electron-doped cuprates.

In this work, we employ resonant inelastic x-ray scattering (RIXS) technique to explore the CDW order in the electron-doped LCCO in a broad temperature and doping concentration range. As commonly realized by doping either electrons or holes to La$_2$CuO$_4$, this system is an excellent platform to reveal the symmetry (or asymmetry) by direct comparison to its hole-doped counterpart, La$_{2-x}$Sr$_{x}$CuO$_4$ (LSCO) and La$_{2-x}$Ba$_{x}$CuO$_4$ (LBCO), where the relationship between CDW and SC has been extensively studied. Superior sensitivity of RIXS allows us to confirm the existence of weak, but clearly short-range CDW with $\xi_{CDW}\sim35$~\AA~in LCCO over a wide doping concentration $0.07\leq x \leq0.17$. Moreover, the intensity of CDW order parameter is vastly suppressed near the optimal doping where the SC phase shows a highest $T_c$ indicating both orders are intimately linked, similar to the hole-doped LSCO \cite{Wen2019,JQLin2020,vonArx2023}. We propose the CDW correlation length as a key parameter to reconcile seemingly inconsistent experimental results on the interplay between SC and CDW orders in different cuprate families. Thus, we establish a unified framework to understand the phenomenology of CDW and its interplay with SC across the electron- and hole-doped sides of cuprate phase diagram.

\section{Results}

\textbf{Charge density waves in LCCO}

The reliability of a doping-dependent study is often hindered by the variation of sample quality associated with the chemical doping. We overcome this issue by choosing the  combinatorial (or ``combi") La$_{2-x}$Ce$_x$CuO$_4$ films, where the Ce doping concentration $x$ changes linearly along one direction of the film (See Methods), to ensure consistent sample quality across a wide doping range \cite{Jin2011,yuan2022scaling}. It is well known that the CDW order in cuprates manifests itself as an enhancement of quasi-elastic scattering intensity at four symmetry-equivalent positions on $H$-$K$ plane, \textbf{Q} = ($\pm\delta$, 0) and (0, $\pm\delta$), in reciprocal space. Here, $\delta$ denotes the wavevector of CDW modulation. To search these CDW reflections, we performed momentum-dependent RIXS (or qRIXS) scans on LCCO films along Cu-O bond direction (or $H$ direction) (Fig.~\ref{fig1}b). The incoming photon energy was tuned to resonate with Cu $L_3$ absorption edge ($\sim$933 eV). Fig.~\ref{fig1}c shows an intensity map obtained from qRIXS scan on LCCO film with $x = 0.07$ at 20 K, plotted as a function of $H$ and energy loss. In addition to dispersive paramagnon excitation ranging from 0.2 to 0.5 eV, a weak enhancement of zero-energy-loss quasielastic scattering intensity appears at \textbf{Q} = (0.25, 0). The corresponding energy-resolved RIXS spectrum at \textbf{Q} = (0.25, 0) is shown in Fig.~\ref{fig1}d and Supplementary Fig.~\blue{1}. The low-energy spectral weights are fitted with a sum of quasielastic (at 0 meV), two phonons ($\sim$40 and $\sim$70 meV), and paramagnon ($\sim$300 meV) contributions in resemblance to the hole-doped La-based cuprates \cite{JQLin2020,Wang2021b,Huang2021,vonArx2023} (See Methods and Supplementary Fig.~\blue{2}). The integrated intensity of the quasielastic peak is plotted as a function of $H$ in Fig.~\ref{fig1}e. An enhancement is clearly identified around $H=0.25$, which can be interpreted as the signature of CDW in LCCO. Similar CDW signal was observed in LCCO single crystal with $x=0.08$~\cite{daSilvaNeto2016}. By fitting with a Gaussian function plus a polynomial background, we extracted the Guassian peak intensity, proportional to the square of CDW order parameter $\Delta_{CDW}$, and its correlation length $\xi_{CDW}$, inferred by the inverse of half-width half-maximum. For LCCO $x=0.07$, $\xi_{CDW}\sim29.5(4)$~\AA. We could not find similar enhancement in the qRIXS scans along $(H,H)$ (or Cu-Cu bonding) direction (Supplementary Fig.~\blue{3}). Fig.~\ref{fig1}f shows RIXS intensity integrated over an energy-loss range [-0.04, 4.0] eV, mimicking energy-integrated RXS experiments. An enhancement of intensity related to CDW is hardly visible near $H=0.25$, as the profile is overwhelmed by $dd$ excitation. It is also absent in the integrated spectral weights from paramagnon (Fig~\ref{fig1}g) and phonon (Fig.~\ref{fig1}h) contributions. Thus, the CDW in LCCO has a quasi-static nature, rather than dynamical with finite excitation energy shown in NCCO \cite{Neto2016}.\\
\newpage

\textbf{Doping and temperature dependence}

 Fig.~\ref{fig2} summarizes the doping dependence of CDW peak measured on LCCO combi films at 15 K. As shown in the background-subtracted data (Fig.~\ref{fig2}a), the CDW peak is broadened and weakened alongside the doping increase from the under-doped ($x=0.07$) to the optimally-doped ($x=0.11$) region where the $T_c$ shows the highest value. The CDW peak disappears at $x=0.11$ and reenters at higher doping levels. Figs.~\ref{fig2}b-d show the doping dependence of CDW peak intensity $I$, wavevector $\textbf{Q}_{CDW}$, and correlation length $\xi_{CDW}$. Most striking observation is that the intensity is vastly suppressed near $x=0.11$ (Fig.~\ref{fig2}b). This behavior has not been observed in the previous doping-dependent study on NCCO \cite{Neto2016}. $\textbf{Q}_{CDW}$ increases monotonically from 0.25 to 0.33, similar to that in NCCO. 
$\xi_{CDW}$ does not show any meaningful doping dependence with the mean value of $\sim$28~\AA, implying the short-range CDW develops in the entire doping range explored in this study.

We further scrutinize the suppression of CDW peak by investigating its temperature dependence. In Fig.~\ref{fig3}a, a series of qRIXS scans at $x=0.11$ LCCO measured under different temperatures are presented. A very weak CDW peak, hardly distinguishable from the background level, is found at 11 K near $H=0.275$. Upon warming up, the CDW peak grows and persists up to 300 K. The temperature dependence of CDW intensity, $\textbf{Q}_{CDW}$, and $\xi_{CDW}$ is summarized in Fig.~\ref{fig3}b-d, respectively. The intensity is enhanced from 11 K to 30 K, but slightly weakened upon further warming, indicating that the CDW suppression is intimately related to the emergence of SC. In contrast to its dramatic effect on the CDW intensity, the temperature does not significantly influence on $Q_{CDW}$ and $\xi_{CDW}$ (Figs.~\ref{fig3}c,d). 
Similar behavior was also observed for $x=0.10$ LCCO (See Supplementary Fig.~\blue{4}).\\

\textbf{Comparison to hole-doped LSCO}

The suppression of CDW peaks inside SC phase have been reported in several hole-doped cuprate materials, evidencing the competition between SC and CDW orders \cite{Comin2015,Tranquada2021,Hayden2023}. 
In LSCO, the coexistence of medium-range CDW ($\xi\sim60$~\AA) near $x=1/8$ doping and short-range CDW ($\xi\sim25$~\AA) in more extended phase diagram was reported by recent RXS experiments \cite{Wen2019,vonArx2023}. Motivated by these works, we compare temperature evolution of CDW peak intensity in underdoped ($x<0.14$) and optimally-doped (or slightly overdoped, $0.14<x<0.20$) region of LSCO reported by different groups employing both resonant and non-resonant x-ray scattering technique in Fig.~\ref{fig4}a and ~\ref{fig4}b, respectively \cite{CroftPRB2014,Christensen2014,Choi2022,Miao2021,Wu2012,Thampy2014,Wen2019,Huang2021,Wang2021,vonArx2023}. In addition, the suppression of CDW peak below $T_c$ is ubiquitously observed for the short-range CDW in optimally-doped region (Fig.~\ref{fig4}b). Our observation of the short-range CDW in LCCO ($\xi\sim35$~\AA) near the optimally-doped $x=0.11$ fits into this category. On the other hands, the medium-range CDW in underdoped LSCO shows distinct behavior: the suppression was commonly found  in non-resonant x-ray scattering works \cite{CroftPRB2014,Christensen2014,Choi2022,Miao2021}, whereas RXS experiments mostly report its enhancement below $T_c$ \cite{Wu2012,Thampy2014,Wen2019,Huang2021,Wang2021,vonArx2023}. This seemingly incompatible dichotomy implies that underdoped LSCO is at proximity of various phases having a comparable total energy. Thus, subtle perturbation in materials or experimental probes may cause a sizable tuning of its ground-state. It should be noted that non-resonant scattering is sensitive to periodic lattice distortion associated with CDW, while RXS directly probes valence electron density associated wtih hybridized Cu 3$d$ - O 2$p$ orbitals. \\

\textbf{Interplay and correlation length}

The reduction of CDW intensity below $T_c$ observed in optimally-doped LSCO and LCCO is not a universal phenomenology shared among different cuprate families. For example, the static long-range CDW in LBCO with $\xi_{CDW}\sim200$~\AA~ does not show any suppression near $x=1/8$ doping where a large dip appears in SC phase   ~\cite{HuckerPRB2011,Thampy2013,Miao2017}. In  prototypical hole-doped Bi-2201, the short-range CDW remains unaffected upon entering SC phase~\cite{Comin2014,Li2020,Choi2024}, in stark contrast to Bi-2212 where the intensity of CDW peak is suppressed below $T_c$~\cite{WSLee2019,Lu2022}. Our CDW results in LCCO show difference to that in electron-doped NCCO in which CDW do not interact with SC phase \cite{daSilvaNeto2018,daSilvaNeto2015,daSilvaNeto2016,Jang2017}. This discrepancy naturally poses the general question of whether the interplay between SC and CDW can be understood within a unified framework or should be treated as  material-specific phenomenology.

We propose that the seemingly diverging experimental results can be reconciled by introducing $\xi_{CDW}$ as a key parameter to describe the interplay between SC and CDW. In Fig.~\ref{fig4}c, $\xi_{CDW}$ of different cuprate materials is classified into four regimes: ($i$) dynamical charge-density fluctuation (CDF, $\xi_{CDW}<\lambda$) where $\lambda$ is the periodicity of density modulation, ($ii$) quasi-static short-range CDW ($\lambda<\xi_{CDW}<50$~\AA), ($iii$) medium-range CDW (50~\AA~$<\xi_{CDW}<100$~\AA), and ($iv$) static long-range CDW order ($\xi_{CDW}>100$~\AA) also known as ``stripe" order. The temperature evolution of CDW intensity and the shape of SC dome in each regime are schematically illustrated in Fig~\ref{fig4}c. The dynamical CDF with ultrashort $\xi_{CDW}$ does not interact with SC and hence is not affected by changing doping or temperature \cite{Arpaia2019}. The reported non-interacting CDW with respect to SC in Bi-2201 \cite{Comin2014} and NCCO \cite{daSilvaNeto2015,daSilvaNeto2016,daSilvaNeto2018,Jang2017} may be explained by their very short correlation lengths ($<20$~\AA) where $\xi_{CDW}$ is comparable to $\lambda$. When $\xi_{CDW}$ exceeds $\lambda$, the weakly correlated CDW starts to interact with SC. CDW is strongly suppressed inside SC phase, as shown in Bi-2212 \cite{WSLee2019} and optimally-doped LSCO \cite{Wen2019}. However, $\xi_{CDW}$ is still not sufficiently long enough to give visible impact on $T_c$. Our results show that LCCO falls into this category. When $\xi_{CDW}$ reaches the long-range static regime ($\xi_{CDW}>100$~\AA), the robust CDW strongly suppresses SC while its intensity monotonically increases upon the temperature decrease as if it does not feel the disruption from the SC order seen in LBCO \cite{Miao2017}.

In between the short-range and long-range CDW regime, there exists the medium-range CDW regime where the impact of CDW on SC becomes stronger than that in the short-range CDW regime and induces a dip in SC phase. In turn the suppression of CDW is somewhat less severe comparing to that in the short-range CDW regime. It may explain the diverging behaviour of CDW observed in the underdoped LSCO (Fig.~\ref{fig4}a). This regime is also of particular interest in terms of putative PDW state that has been recently proposed as a novel form of intertwined orders \cite{Fradkin2010,Agterberg2020}. Despite the evidence provided by scanning tunneling microscopy \cite{Hamidian2016,Edkins2019,Du2020}, its direct observation in bulk materials remain elusive. From ``subdominant order" perspective \cite{JSLee2023}, the medium-range CDW regime where SC, CDW, and SDW orders are moderately correlated and coexist without vanishing each other may be a good candidate to search for putative PDW order.

\section{Discussion}



Fig.~\ref{fig5} depicts the phase diagrams of LCCO and LSCO. Our results add the presence of short-range CDW over a wide temperature and doping range and its suppression below $T_c$ to the list of phenomenology universally present in both sides of phase diagram. This remarkable universality implies that the short-range CDWs in both sides may come from a common physical origin. Their characteristics do not change qualitatively across the critical doping for Fermi surface reconstruction (FSR) in both LCCO and LSCO \cite{Lin2020}, likely because $\xi_{CDW}$ is too short to trigger FSR. This result supports that the short-range CDW is commonly driven by strong electronic correlations rather than by weak-coupling Fermi surface nesting in both sides of La-based cuprates.

However, the phenomenology of CDW has also asymmetric aspects: the medium-range CDW appears near $x=1/8$ hole doping and interacts with SC in a distinct way. This asymmetry may be linked to the formation of intertwined spin-charge ``stripe" order, where segregation of doped holes lowers the total energy and stabilizes spin and charge density modulations with a mutual commensuration $\delta_{CDW}=2\delta_{SDW}$ \cite{Tranquada2021}. The interplay between SC and CDW is often described by the framework of phenomenological Landau theory \cite{Zachar1998,Nie2017,ZDYu2017}. The quadratic term in Landau free energy coupling of two competing order parameters well captures the suppressed CDW intensity below $T_c$. When a cubic coupling term between SDW and CDW that vanishes unless $\delta_{CDW}=2\delta_{SDW}$ is introduced to the model, it could modify the delicate energy balance among different orders and gives rise to more complex interplay such as cooperative relationship between SC, CDW and SDW order~\cite{Wen2019}. In LCCO, on the other hands, the absence of SDW simplifies the picture -- competition between SC and CDW order. Recent quantum Monte Carlo calculation revealed the above asymmetry \cite{Mai2022}. 

Alternatively, the suppression of CDW intensity can be understood from the viewpoint of quantum phase transition. Recent RIXS works reported substantial enhancement of CDF with finite excitation energy $\Delta\sim20$ meV upon cooling below $T_c$ accompanying suppressed CDW intensity  \cite{WSLee2019,Huang2021,Lu2022,Arpaia2023}. The authors proposed a concept of enhanced quantum fluctuation near quantum critical point (QCP). When the system enters SC state (or approaches QCP), dissipation coupled to electrons in Fermi surface is significantly suppressed and quantum fluctuations becomes more severe than those above $T_c$. This scenario provides new insights on the relationship between SC and CDW order beyond a simple competing picture based on Landau theory \cite{WSLee2019}. In electron-doped LCCO, we could not identify the enhanced inelastic spectral weights associated with the suppression of CDW order near the optimal doping. However, this result does not rule out the dissipation-driven QCP scenario. $\Delta_{SC}$ of LCCO, expected to be similar to that of NCCO \cite{Sato2001} ($\sim10$ meV), is much smaller than in Bi-2212 ($\Delta_{SC}\sim80$ meV)~\cite{Hashimoto2014}. Thus, the SC order may not be strong enough to enhance quantum fluctuations in the energy scale accessible with the resolution in this work.

In summary, our high-resolution RIXS experiments on LCCO films reveals ubiquitous existence of short-range CDW over broad temperature and doping ranges. The intensity of CDW peak was significantly suppressed below $T_c$ near the optimal doping $x\sim0.11$. This observation confirms that both CDW and SC orders are intertwined in both sides of La-based cuprate phase diagram. By suggesting the CDW correlation length as a key parameter, we establish more unified picture to describe the interplay between SC and CDW in different cuprate materials.\\ 
\\





\indent\textbf{METHODS}\\

\begin{figure*}[t]
\center{\includegraphics[width=0.96\textwidth]{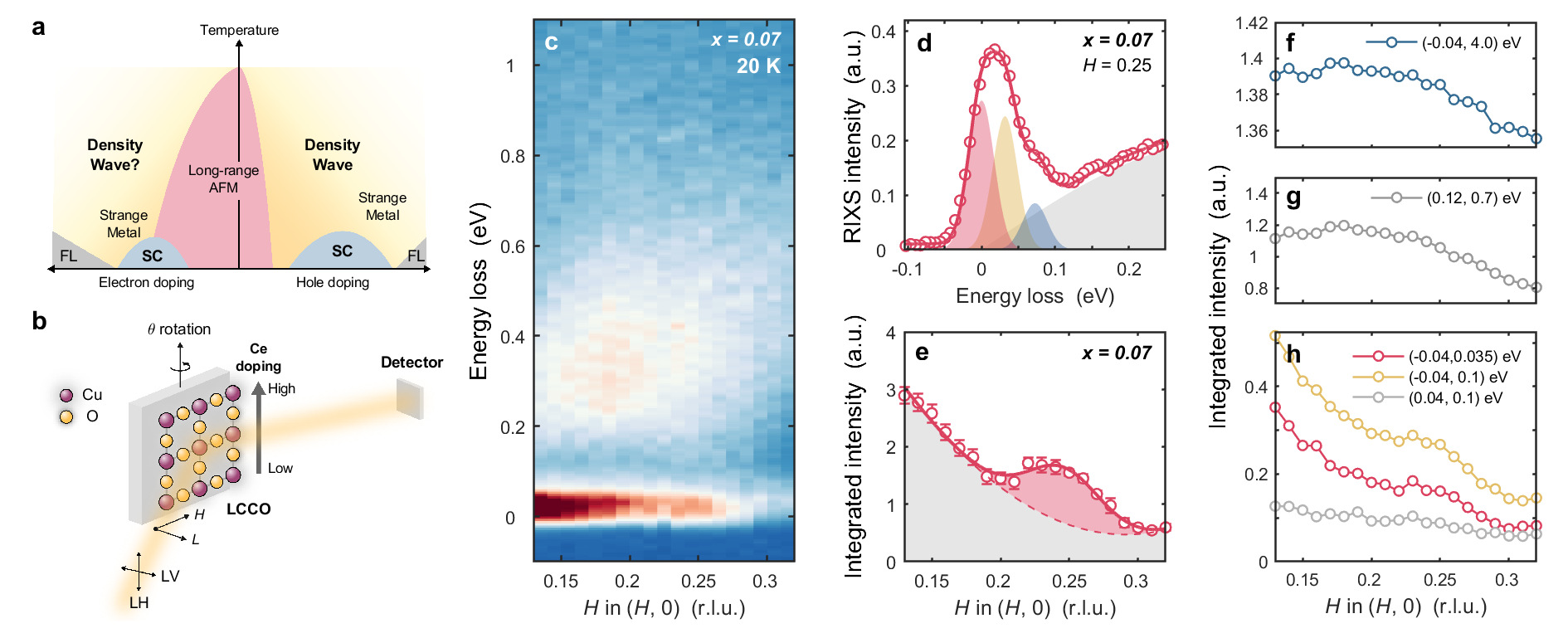}}
\caption{\textbf{Charge density waves in electron-doped LCCO with Ce doping concentration $x=0.07$.} (\textbf{a}) Schematic phase diagram of cuprates. SC and FL refers superconducting and Fermi liquid phase, respectively. (\textbf{b}) RIXS experiments on LCCO thin film with a linear gradient of Ce doping concentration. (\textbf{c}) Momentum-dependent Cu $L$-edge RIXS map measured at 20 K for doping concentration $x=0.07$. (\textbf{d}) RIXS spectrum of LCCO $x=0.07$ at $H=0.25$. The red, yellow, blue and grey shaded region represents contribution from quasielastic, two phonons, and paramagnon excitation, respectively. (\textbf{e}) Momentum profile of quasielastic contribution integrated over an energy-loss range of (-0.35, 0.35) eV along the Cu-O bond direction. The solid line indicates a Gaussian function (red) with a polynomial background (grey). RIXS intensity integrated over (\textbf{f}) (-0.04, 4.0) eV, (\textbf{g}) (0.12, 0.7) eV, (\textbf{h}) (-0.04, 0.035) eV (red circles), (-0.04, 0.1) eV (yellow circles), and (0.04, 0.1) eV (light grey circles) is plotted as a function of $H$. 
}	
\label{fig1}
\end{figure*}

\begin{figure}[t]
\center{\includegraphics[width=0.8\textwidth]{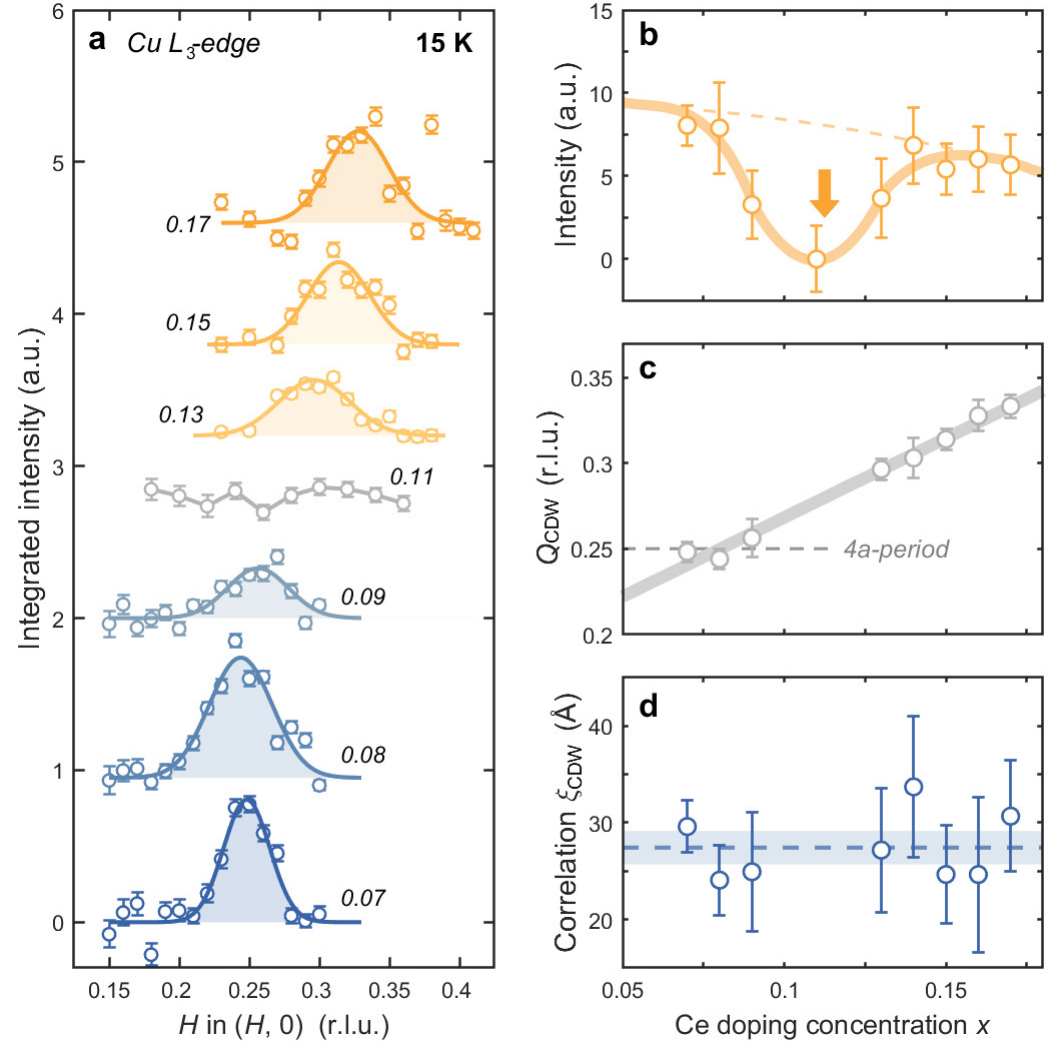}}
\caption{\textbf{Doping dependence of charge density waves in electron-doped LCCO.} (\textbf{a}) qRIXS scan profiles across CDW in LCCO at different doping concentration as indicated. Polynomial background was subtracted from each profile after Gaussian fitting (See Method). (\textbf{b}) Peak intensity, (\textbf{c}) wavevector $Q_{CDW}$ and (\textbf{d}) correlation length $\xi_{CDW}$ plotted as a function of doping concentration. All data shown in this figure were obtained from ``Combi 1" LCCO film, except for $x=0.17$ data from ``Combi 2" LCCO film.
}	
\label{fig2}
\end{figure}

\begin{figure}[t]
\center{\includegraphics[width=0.8\textwidth]{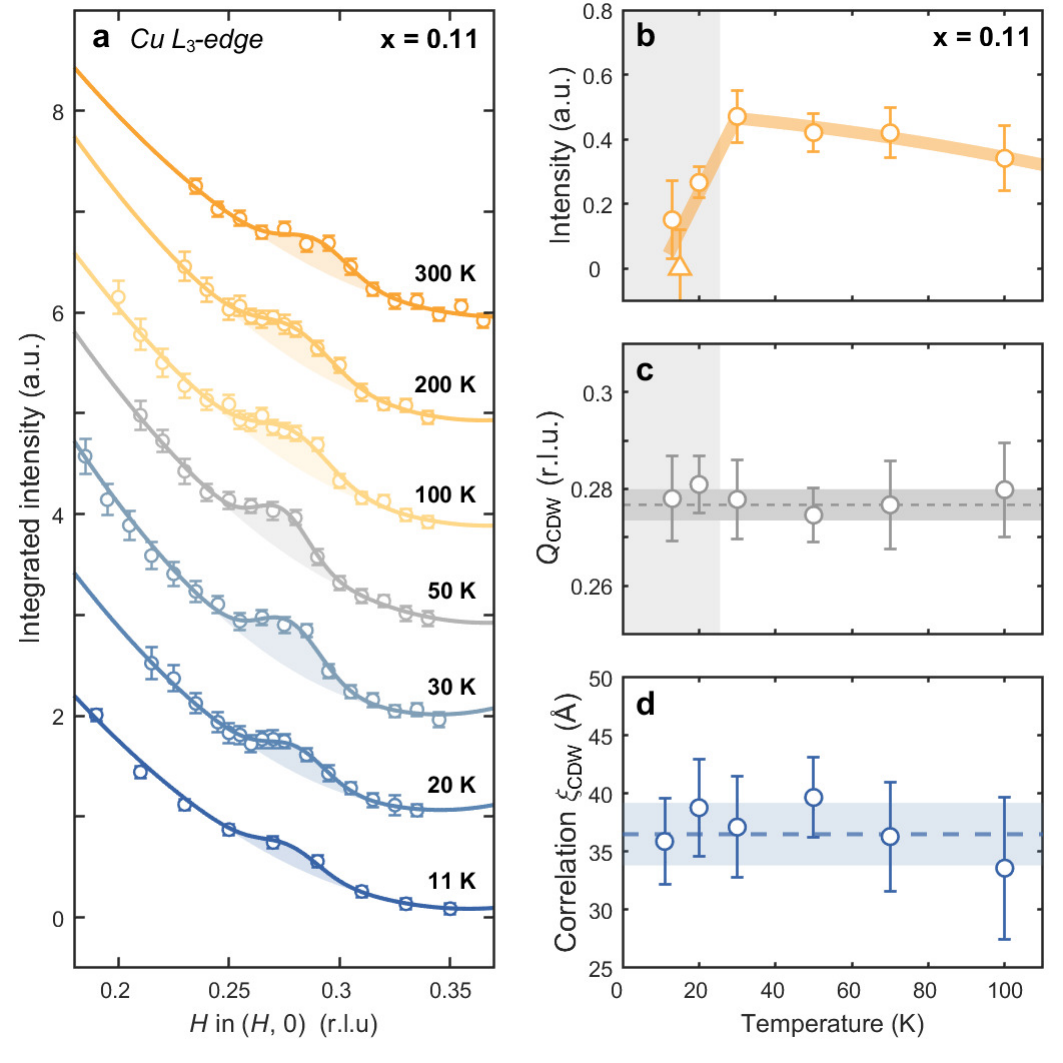}}
\caption{\textbf{Temperature evolution of CDW order.} (\textbf{a}) Raw momentum-dependent CDW profiles measured at different temperatures as indicated. The solid lines indicate the fits to a Gaussian peak with a polynomial background for each profile. (\textbf{b}) Peak intensity, (\textbf{c}) wavevector $Q_{CDW}$ and (\textbf{d}) correlation length $\xi_{CDW}$ plotted as a function of temperature. The grey-colored area is the region where SC phase appears. The dashed line indicates $T_c$. All temperature-dependence data were obtained from ``Combi 2" LCCO film, except for the open triangle data in the panel (b).
}	
\label{fig3}
\end{figure}

\begin{figure*}[t]
\center{\includegraphics[width=1\textwidth]{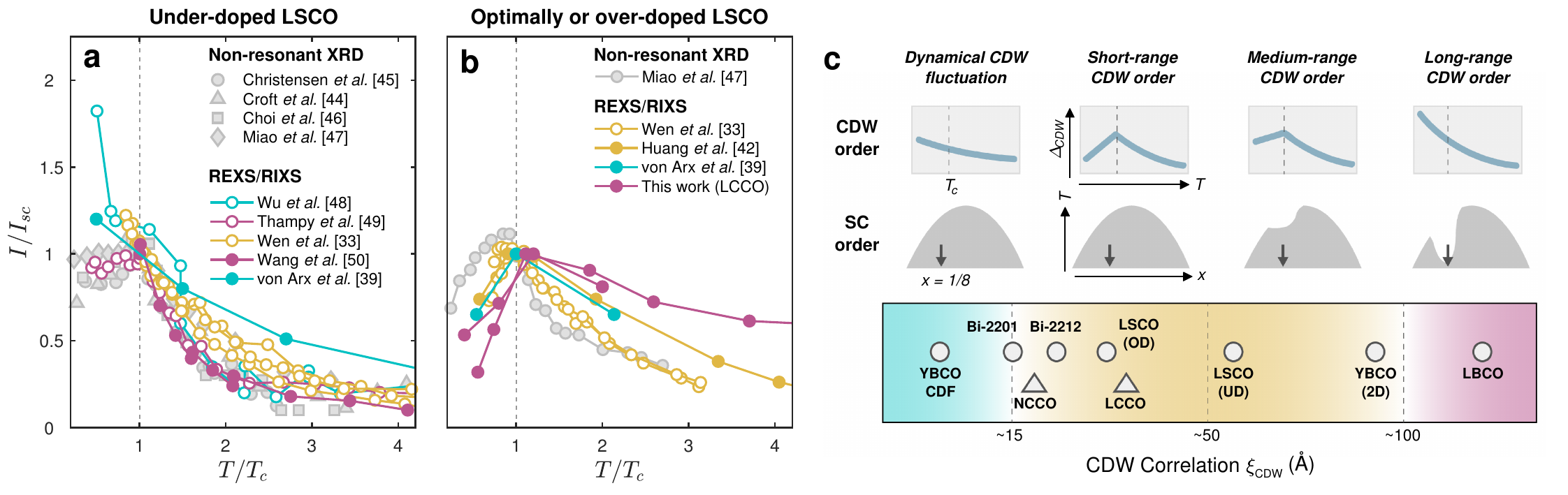}}
\caption{\textbf{Interplay between superconductivity and CDW order.} Comparison of CDW peak intensity in (\textbf{a}) underdoped LSCO ($x<0.14$) and (\textbf{b}) optimally- and over-doped LSCO ($0.14<x<0.20$), reported by different groups employing non-resonant \cite{CroftPRB2014,Christensen2014,Choi2022,Miao2021} and resonant x-ray scattering techniques \cite{Wu2012,Thampy2014,Wen2019,Huang2021,Wang2021,vonArx2023}. x and y axis represents the temperature and peak intensity scaled by $T_c$ and $I_{sc}$, the CDW peak intensity at $T=T_c$, respectively. The grey dashed line indicates $T_c$ at each dataset. (\textbf{c}) Graphical illustration of the interplay between SC and CDW orders in four different regimes of $\xi_{CDW}$ (See main text). In the upper panels, typical temperature dependence of CDW order parameter is illustrated in four different CDW regimes. The middle panels schematically represents the shape of SC dome that corresponds to each CDW regime. The vertical and horizontal axis indicates $T_c$ and doping concentration, respectively. The dashed lines and arrows indicate $T_c$ and $x=1/8$ doping in the upper and middle panels, respectively. Representative $\xi_{CDW}$ measured at low temperature for different cuprate materials including YBCO~\cite{Ghiringhelli2012,Chang2012,Choi2020,Arpaia2019}, Bi-2201~\cite{Comin2014,Li2020,Choi2024}, Bi-2212~\cite{WSLee2019,Lu2022}, LSCO~\cite{CroftPRB2014,Christensen2014,Choi2022,Miao2021,Wu2012,Thampy2014,Wen2019,Huang2021,Wang2021,vonArx2023}, LBCO~\cite{HuckerPRB2011,Thampy2013,Miao2017}, NCCO~\cite{daSilvaNeto2015,daSilvaNeto2016,Jang2017} and LCCO (this work) reported by various x-ray scattering experiments are summarized in the lower panel.
}
\label{fig4}
\end{figure*}

\begin{figure*}
\center{\includegraphics[width=0.92\textwidth]{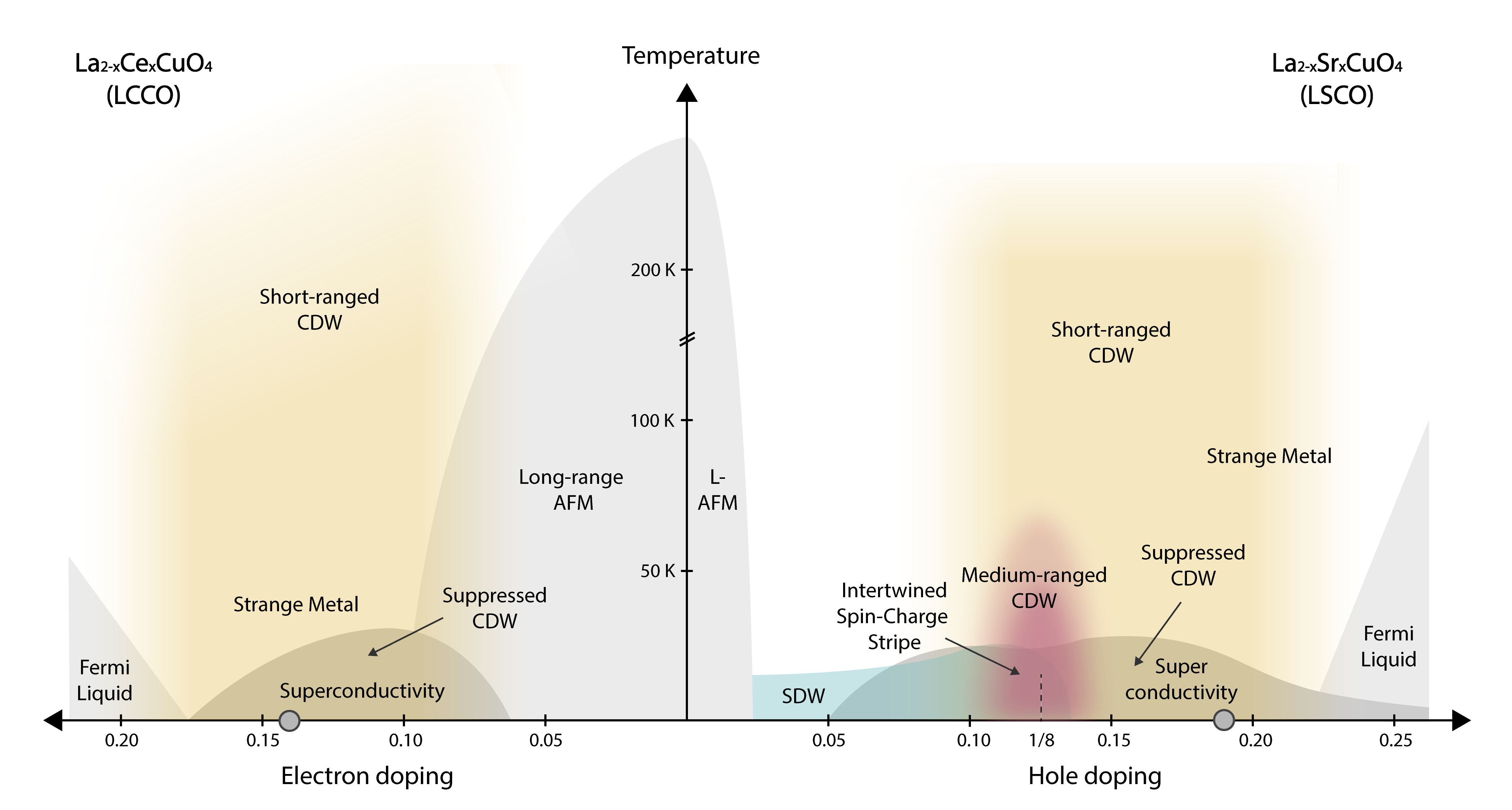}}
\caption{\textbf{Phase diagram of Charge-density-wave phenomenology in La-based cuprates.} A short-range CDW with $\xi_{CDW}\sim30$~\AA~appears in both electron- (left panel) and hole-doped (right) sides of phase diagram over a wide range of temperature and doping concentration. On the other hands, the presence of a medium-ranged CDW with $\xi_{CDW}\sim60$~\AA is found to be specific to hole-doped side (right panel). This medium-range CDW intimately interacts with superconductivity by forming an intertwined spin-charge stripe order. The regions occupied by other phases are adopted from Ref.~\cite{Wen2019,Armitage2010,Greene2020,vonArx2023,Miao2021}. Grey circles on x-axis indicate $x_{FSR}$, the critical doping concentration for Fermi surface reconstruction.
}
\label{fig5}
\end{figure*}

\textbf{Sample preparation}\\
High-quality La$_{2-x}$Ce$_{x}$CuO$_{4}$ (LCCO) films were grown on SrTiO$_3$ substrates via the laser molecular beam epitaxy (LMBE) technique with $\sim$160-nm thickness. The samples are stacked such that the $c$-axis is normal to the surface. Most of the experimental data presented here were collected from LCCO combinatorial (or ``combi") film samples in which the Ce concentration $x$ continuously varies with a linear relation with the spatial location of the samples. For this work, two samples ``Combi 1" and ``Combi 2" were fabricated by the continuous moving mask technique. More information on this method is available in Ref.~\cite{Yu2017}. Each sample possesses a continuous gradient of chemical composition $x$ ranging from 0.07 to 0.15 (Combi 1) and from 0.10 to 0.19 (Combi 2) along the vertical $z$ direction, which covers a broad region in the phase diagram spanning the entire superconducting dome \cite{Jin2011}. The ``Combi 1" sample was mostly used for the doping-dependent study in Fig.~\ref{fig2} except for the data with the highest doping $x=0.17$. The temperature-dependent study in Fig.~\ref{fig3} was solely obtained from the ``Combi 2" sample. The $c$-axis lattice constant and superconducting transition temperature $T_c$ measured at different $z$ positions were consistent with results from single-doping LCCO films \cite{Sawa2002}, confirming the superb quality of combi samples. \\
\\
\textbf{High-resolution RIXS experiments}\\
We performed high-resolution RIXS experiments on LCCO films at the I21 beamline of Diamond Light Source, United Kingdom \cite{KJZhou2022}. The LCCO possesses a tetragonal symmetry (\textit{I4/mmm}) with $a=b\sim4.01$~\AA~and $c$ linearly varying from 12.47~\AA~(x = 0.07) to 12.40~\AA~(x = 0.17). Throughout the manuscript, the wavevector \textbf{Q} = ($H$, $K$, $L$) is defined as ($2\pi/a$, $2\pi/b$, $2\pi/c$) employing a pseudo-tetragonal notation in reciprocal lattice unit (r.l.u.). The Ce doping concentration $x$ varies from 0.07 (0.09) to 0.15 (0.17) for Combi 1 (Combi 2) sample along the $z$ position of the samples, perpendicular to the horizontal scattering plane. As illustrated in Fig.~\ref{fig1}\blue{b}, the samples were mounted on a cryogenic manipulator with six degrees of freedom such that the crystallographic $a$- and $c$-axis (or equivalently $b$-c) are lying on the horizontal scattering plane. Prior to RIXS experiments, x-ray absorption spectroscopy (XAS) was performed on LCCO films to tune the incident photon energy to Cu $L_{3}$ absorption edge ($\sim$933 eV). Both fluorescence and electron yields were measured. Linear vertical (LV, or $\sigma$) polarization was used unless otherwise stated. We performed momentum-dependent RIXS scans along the Cu-O bond direction (defined here as $H$ direction) by changing the sample rotation angle $\theta$, while the scattering angle $2\theta$ was fixed to 154$^\circ$. A paraboloidal collecting mirror was implemented to maximize the scattered x-ray beam throughput. Finally, we obtained the resolution of 37.4 meV at Cu $L_3$-edge. We controlled the sample temperature with heaters mounted on the Liquid He flow cryostat in a range of 11$\sim$300 K.  \\
\\
\textbf{Data fitting}: \\
The collected RIXS spectra were first normalized by the flux of incident photons and then corrected for the self-absorption effect. To disentangle unwanted inelastic contributions, the low-energy part of the RIXS spectra was fitted with a sum of quasielastic scattering, phonon and paramagnon excitations, and polynomial backgrounds (See Supplementary Fig.~\blue{2}). We fit the quasielastic peak and phonon excitations using Gaussian function, while the paramagnon excitation is best described by the shape of a damped harmonic oscillator response \cite{Monney2016}. The FWHM of Gaussian functions was fixed to the energy resolution. The RIXS spectra were shifted such that the position of the quasielastic peak is calibrated to the zero energy loss. To analyze the momentum dependence of CDW, we extracted the integrated intensities of the quasielastic peak from individual spectra and plotted them as a function of $H$. This $H$-dependent profile was subsequently fitted with a Gaussian peak with a polynomial background. The peak amplitude is proportional to the square root of CDW order paramter $\Delta$, \textit{i.e.,} $I\sim\sqrt{\Delta}$. Its in-plane correlation length is defined as $\xi\sim 2/\Gamma$ where $\Gamma$ is a FWHM of the Gaussian peak.   \\
\\
\textbf{Acknowledgements:}\\
We thank Wei-Sheng Lee, Masafumi Horio, and Qisi Wang for fruitful discussion. This work was primarily supported by user research program of Diamond Light Source, Ltd. through beamtime proposal MM27872. J.C. acknowledge financial support from the National Research Foundation of Korea (NRF) funded by the Korean government (MSIT) through Sejong Science Fellowship (Grant No. RS-2023-00252768).\\ 
\\
\\

\normalsize

\bibliography{lcco_ref}

\newpage

\newpage

\newpage

\newpage

\newpage

 \end{document}